\documentclass[twocolumn]{revtex4-1}
\usepackage{graphics}
\usepackage{graphicx}
\usepackage{amsmath}
\usepackage{amssymb}
\usepackage{bm}         

\renewcommand{\eqref}[1]{(\ref{eq:#1})}

\newcommand{\figref}[1]{Fig.~\ref{fig:#1}}
\newcommand{\Figref}[1]{Figure~\ref{fig:#1}}

\newcommand{\citeasnoun}[1]{Ref.~\citenum{#1}}

\begin{document}

\title{Experimental Demonstration of Quasi-Resonant Absorption \\ in Silicon Thin Films for Enhanced Solar Light Trapping}

\author{Ardavan Oskooi}
\email{oskooi@qoe.kuee.kyoto-u.ac.jp}
\author{Menaka De Zoysa}
\author{Kenji Ishizaki}
\author{Susumu Noda}
\affiliation{Department of Electronic Science \& Engineering, Kyoto University, Kyoto 615-8510, Japan}

\begin{abstract}
We experimentally demonstrate that the addition of partial lattice
disorder to a thin-film micro-crystalline silicon photonic crystal
results in the controlled spectral broadening of its absorption peaks
to form quasi resonances; increasing light trapping over a wide
bandwidth while also reducing sensitivity to the angle of incident
radiation. Accurate computational simulations are used to design the
active-layer photonic crystal so as to maximize the number of its
absorption resonances over the broadband interval where
micro-crystalline silicon is weakly absorbing before lattice disorder
augmented with fabrication-induced imperfections are applied to
further boost performance. Such a design strategy may find practical
use for increasing the efficiency of thin-film silicon photovoltaics.
\end{abstract}


\maketitle

\noindent A key component of a thin-film silicon photovoltaic (PV)
cell is a light-trapping structure that provides enhanced absorption
of solar radiation over a broad range of frequencies and incident
angles~\cite{Poortmans06}. While there have been numerous proposals of
nanostructured designs exploiting resonant wave effects of photons for
this purpose~\cite{Pillai07,Garnett10,Han10,Sheng11,Hsu12,Sai12}, most
have been limited by inherent delicate-interference effects to a
restricted set of operating conditions involving narrow bandwidths,
select polarizations or a small angular cone. We recently introduced a
new mechanism for photon absorption in thin films based on the notion
of \emph{quasi} resonances which combine the large absorption of
impedance-matched resonances with the broadband and robust
characteristics of disordered systems~\cite{Oskooi12}. Furthermore we
showed that such an approach can significantly exceed the Lambertian
light-trapping limit - the maximum possible for optically-thick wafer
cells employing random scattering of light rays - over a wide range of
conditions~\cite{Oskooi13}. In this letter, we experimentally
demonstrate using conventional semiconductor nanofabrication processes
potentially applicable to large-scale thin-film PV manufacturing how
quasi resonances can be used to increase the light-trapping
performance of a micro-crystalline silicon ($\mu$c-Si:H) thin film by
applying partial lattice disorder to an active-layer photonic crystal
(PC) in order to precisely control the spectral broadening of its
resonant-absorption peaks. We employ computational design based on
accurate finite-difference time-domain (FDTD)
simulations~\cite{Oskooi10} to structure a thin-film PC slab to
contain as many absorption resonances as possible over a broad
bandwidth where $\mu$c-Si:H is weakly absorbing and also given the
consistency of the numerical and experimental results use such
simulations to investigate absorption in the rest of the device which
is precluded by an experimental approach alone. Finally, we show how
the presence of unintentional fabrication-induced imperfections can be
harnessed together with intentional lattice disorder to further
enhance light trapping over a broad bandwidth and angular range.

\begin{figure}
{\centering
  \includegraphics[width=1.0\columnwidth]{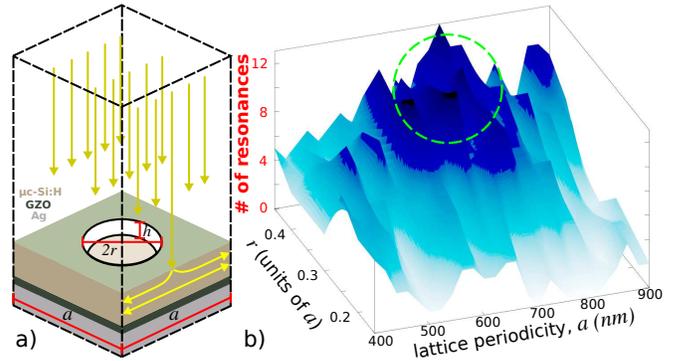} \par}
\caption{The $\mu$c-Si:H PC active layer is computationally designed
  to maximize the number of its absorption resonances which is
  correlated with its light-trapping efficiency~\cite{Oskooi13}. (a)
  Schematic of the unit cell for the thin-film PC device used in the
  FDTD simulations to compute the absorptivity spectra at normal
  incidence. The PC consists of a square lattice of holes in the
  $\mu$c-Si:H layer with periodicity $a$, hole radius $r$ and height
  $h$. (b) 3D contour plot of the number of resonances - obtained from
  the simulated absorptivity spectra by using a constant
  peak-identification threshold criteria of 80\% absorptivity with no
  other constraints - plotted over the 2D parameter space of PC
  lattice periodicity $a$ and hole radius $r$ (hole height $h$ is
  fixed at 180nm). The region containing the largest values is circled
  in green.}
\label{fig:figure1}
\end{figure}

A substrate-type thin-film $\mu$c-Si:H PV cell with light-trapping
capability is typically fabricated by depositing the active layer on a
textured back reflector such that the pattern is then transferred
throughout the entire device~\cite{Python09,Sai12}. This specific
design however has two principal disadvantages: firstly, patterning of
the metallic layer may enhance unwanted plasmonic
losses~\cite{Palanchoke12} and secondly, the $\mu$c-Si:H layer must be
grown on a non-uniform surface which due to the onset of more grain
boundaries and defects during film growth act as recombination centers
degrading its charge-transport properties~\cite{Python09}. As a
result, in this work we take a different approach and form the
light-trapping structure directly within only the active layer
itself. Our device, a precursor to a fully-functional PV cell, is
fabricated on top of a base substrate using conventional thin-film
deposition tools as follows. First, a back reflector consisting of a
270nm silver (Ag) film is placed on the substrate via electron-beam
physical-vapor deposition. Next, a thin 40nm layer of gallium-doped
zinc oxide (GZO) with carrier density and mobility of
5x10$^{19}$cm$^{-3}$ and 18cm$^2$/(Vs) is sputtered on top followed by
growth of the main active layer: 510nm of intrinsic $\mu$c-Si:H via
plasma-enhanced chemical vapor deposition (PECVD) with hydrogen
(H$_2$) and silane (SiH$_4$) feed-in gases. The crystallinity fraction
of the $\mu$c-Si:H film is found using Raman spectroscopy to be 59\%
by fitting it to a combination of the known spectra of amorphous and
crystalline silicon~\cite{Vallat06}. Then, electron-beam lithography
is used to define the two-dimensional PC pattern over an area of
2107x2107 periods of the square lattice (approximately 1.5mm$^2$) in
the spun-on positive resist which is followed by inductively-coupled
plasma (ICP) reactive-ion etching (RIE) to create arrays of
cylindrically-shaped holes on the top surface. Lastly, the resist
residue is removed by soaking the samples in trichloroethylene
followed by O$_2$ plasma ashing. Note that while this device is
sufficient for investigating the optical-absorption characteristics of
the active $\mu$c-Si:H layer, the focus of this work, it can be
readily extended to a complete PV cell through an additional three
processing steps: growing a $p$-doped $\mu$c-Si:H layer on top of a
hole-etched intrinsic layer which sits above an $n$-doped layer to
form an \emph{n-i-p} junction, sputtering another layer of transparent
conductive oxide (e.g. ITO) with thickness of roughly 70nm on top of
this $\mu$c-Si:H layer and finally depositing a finger-grid of silver
electrodes on the front surface. The electron-beam lithography and dry
plasma etching used here to create a prototype design can be easily
replaced by standard photo-lithography followed by a wet etch which is
more practical for large-area fabrication~\cite{Sai12}. Etching holes
into the $\mu$c-Si:H film is likely to introduce structural defects
though confined to just the top surface; however if the holes can be
made both sufficiently small and smooth while still retaining features
of the Bloch modes the impact of such film-quality degradation may be
mitigated while still providing many more degrees of design freedom in
the choice of lattice for light trapping.

\begin{figure}
{\centering
  \includegraphics[width=1.0\columnwidth]{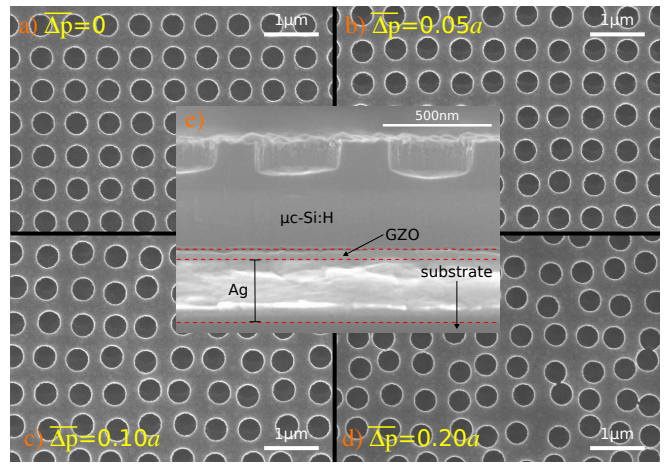} \par}
\caption{Scanning electron microscopy (SEM) images of the thin-film
  light-trapping device showing four partially-disordered PC lattices
  in (a)-(d): the hole positions of the perfect lattice have been
  perturbed by an amount $\Delta p$ chosen randomly from a uniform
  distribution of values between 0 and $\overline{\Delta p}$ for both
  orthogonal in-plane directions. The disorder quantity is given in
  units of the square-lattice parameter, $a$. (e) SEM image of the
  device cross section showing all three individual layers: silver
  (Ag), gallium-doped zinc oxide (GZO) and $\mu$c-Si:H. Note the uneven
  profile of the etched holes in the $\mu$c-Si:H which will cause a slight
  degree of intrinsic broadening in the film's absorption peaks.}
\label{fig:figure2}
\end{figure}

The thin-film PC structure which maximizes the light-trapping
efficiency, a figure of merit we defined previously in
~\citeasnoun{Oskooi12}, can be obtained from a two-part design
strategy~\cite{Oskooi13}: 1) maximize the \emph{number of
  resonant-absorption modes} and 2) apply a partial amount of disorder
to boost broadband light trapping and robustness. While we have
already shown that there exists a close correlation between the number
of absorption resonances and the light-trapping
efficiency~\cite{Oskooi13}, it is worthwhile to reiterate that the
design process is facilitated by the former due to both its direct
connection to quasi resonances and better suitability as an objective
function for topology optimization where a gradient-free search of a
large parameter space is necessary. This is the case even though the
latter which is proportional to the short-circuit current is a
more-explicit measure of PV cell performance. For the first step, we
use the open-source FDTD software tool Meep~\cite{Oskooi10} to
exhaustively explore the PC design space consisting of just two
parameters in its unit cell as shown in the schematic of
~\figref{figure1}a: the lattice periodicity and hole radius (the hole
height is fixed at 180nm; a somewhat arbitrary value that nonetheless
gives rise to meaningful results). In order to ensure that the
absorptivity spectra of our computational design agrees well with its
experimental realization, we need to accurately incorporate the full
complex-refractive index profile of $\mu$c-Si:H over a broad bandwidth
into the FDTD simulations. Here we focus on the wavelength interval
spanning 700nm to 1000nm where the absorption coefficient of
$\mu$c-Si:H is small (less than approximately 1000cm$^{-1}$ and
decreasing~\cite{Vallat06}) and thus the need for a light-trapping
design is most relevant. As it turns out, in this specific regime the
complex-refractive index profile of $\mu$c-Si:H is very similar to
that of crystalline silicon (c-Si)~\cite{Vetterl00,Vallat06} allowing
us to make use of the extensively-compiled data of the
latter~\cite{Green08}. We therefore perform a nonlinear fit of the
c-Si experimental data to a slightly-modified Drude-Lorentzian
polarizability term used in the FDTD simulations~\cite{Oskooi13} and
obtain a close match in both the real part of the refractive index and
the absorption coefficient. For even greater accuracy, we also
incorporate into the simulations the material losses of GZO used in
our actual device by first measuring its absorptivity in a separate
experiment and then from this computing its absorption coefficient
using a transfer-matrix approach~\cite{Yeh88}. We will demonstrate
later on that though GZO's absorption coefficient may be large
relative to that of $\mu$c-Si:H (particularly at long wavelengths),
nevertheless its absorption as a fraction of the total in the overall
device is insignificant given that it is so much thinner and placed
behind the $\mu$c-Si:H layer. ~\Figref{figure1}b shows a contour plot
of the number of absorption resonances - obtained from the simulated
absorptivity spectra at normal incidence where the peak-identification
threshold criteria is taken to be a constant 80\% absorptivity at all
wavelengths with no constraints on the peaks' width or spacing
relative to other peaks - as a function of the lattice periodicity and
radius. The dotted green circle denotes the region where the number of
peaks is largest and from which we select exemplary PC parameters for
fabrication: $a$=650nm and $r$=208nm.

\begin{figure}
{\centering
  \includegraphics[width=1.0\columnwidth]{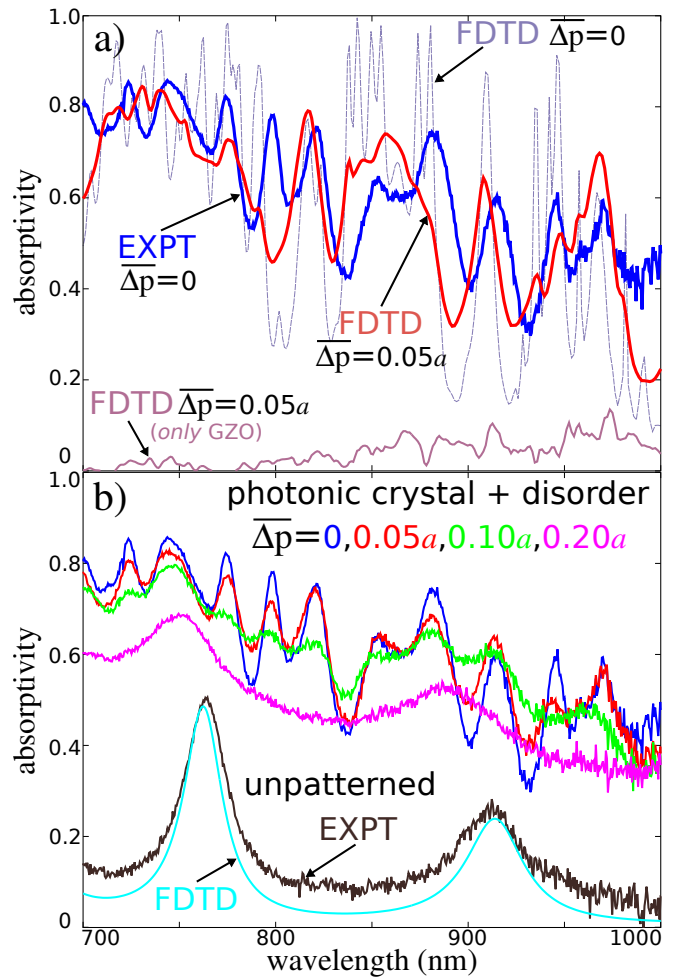} \par}
\caption{(a) Comparison of the absorptivity spectra of the $\mu$c-Si:H
  thin-film PC device for $\mathcal{S}$-polarized light at an angle of
  incidence of 10$^\circ$. The experimental data for the device with
  no disorder ($\overline{\Delta p}$=0, in blue) measured using an
  integrating sphere shows good agreement with the FDTD calculation
  with small disorder ($\overline{\Delta p}$=0.05$a$, in red)
  indicating the presence of intrinsic imperfections. The absorption
  within just the GZO layer in gray as computed by simulations is only
  a small fraction of the overall total. (b) Experimental absorptivity
  data under the same measurement conditions demonstrating gradual
  broadening of the absorption peaks with increasing disorder in the
  PC device. The unpatterned $\mu$c-Si:H film (in black) is shown for
  reference with the corresponding FDTD calculation (in cyan); the
  minor discrepancy arises from the incomplete collection of light
  within the integrating sphere which appears as a constant background
  absorption.}
\label{fig:figure3}
\end{figure}

The fabricated PC device along with three of its partially-disordered
variants where the position of the holes in the perfect lattice has
been perturbed by an amount $\Delta p$ chosen randomly from a uniform
distribution of values between 0 and $\overline{\Delta p}$ for both
orthogonal in-plane directions ($\overline{\Delta p}$=0.05$a$,
0.10$a$, 0.20$a$) is shown in ~\figref{figure2}a-d. Some slight
non-uniformity is evident in the hole shapes which will naturally tend
to broaden the sharp absorption peaks found in the otherwise perfect
case as will be apparent soon. A cross section of the entire device is
shown in ~\figref{figure2}e where the three individual layers can be
seen as well as the profile of the etched holes, with its somewhat
uneven morphology which also contributes to the intrinsic
imperfections affecting device performance. The presence of surface
roughness on the $\mu$c-Si:H film with feature sizes of less than 20nm
does not significantly affect the absorption properties of the device
unlike the irregularities associated with the holes which we confirmed
separately with simulations; this is to be expected since the latter
form the primary scattering objects of the lattice and thus more
strongly influence the delicate-interference effects underlying the
coherent nature of the guided resonant modes.

\begin{figure}
{\centering
  \includegraphics[width=1.0\columnwidth]{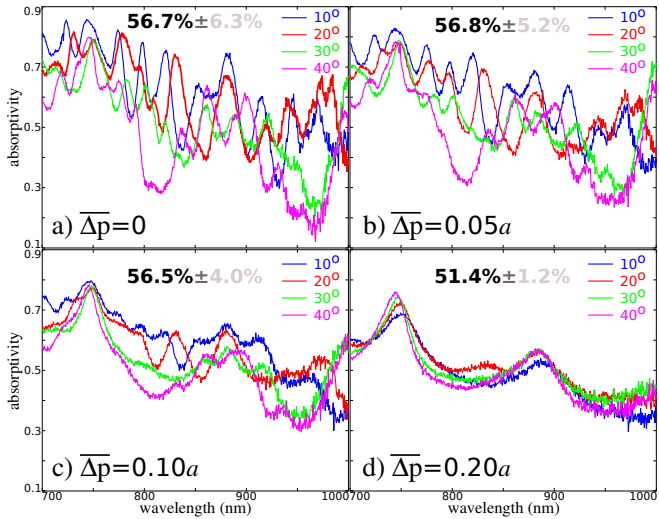} \par}
\caption{Experimental absorptivity data of four partially-disordered
  PC devices ($\overline{\Delta p}$=0, 0.05$a$, 0.10$a$, 0.20$a$) in
  (a)-(d) for $\mathcal{S}$-polarized light at four angles of
  incidence - 10$^\circ$, 20$^\circ$, 30$^\circ$ and 40$^\circ$ -
  where the average light-trapping efficiency (as calculated over
  these four angles) and corresponding standard deviation are
  indicated. The device with nearly maximum light-trapping efficiency
  in (c) with $\overline{\Delta p}$=0.10$a$ is nearly a 10\%
  efficiency improvement above and nearly twice as robust as the ideal
  non-perturbed design.}
\label{fig:figure4}
\end{figure}
 
Next we measure the absorptivity spectra via 1-reflectivity of the
unperturbed ($\overline{\Delta p}$=0) sample in an integrating sphere
connected to a spectrometer and multi-channel Si detector using a
tungsten halogen lamp source with a 610nm low-pass filter for
$\mathcal{S}$-polarized light at an angle of incidence of
10$^\circ$. The results shown in ~\figref{figure3}a (in blue) display
traces of numerous absorption resonances but all have noticeably
undergone a minor degree of broadening due to the non-uniformities
inherent in the holes' morphology as discussed previously. This
becomes evident when these experimental results are compared with
those from two FDTD simulations (computed using identical
incident-light conditions as the experimental measurements): one for
$\overline{\Delta p}$=0 and the other for $\overline{\Delta
  p}$=0.05$a$ where good agreement is apparent between the fabricated
sample with $\overline{\Delta p}$=0 and simulated one with
$\overline{\Delta p}$=0.05$a$ (here three separate simulations are
made using a 10x10 supercell and the absorptivity spectra is averaged
due to the random nature of the design). This therefore suggests that
from the perspective of light trapping the amount of intrinsic
imperfections present within the fabricated sample is approximately
equivalent to a positional disorder of $\overline{\Delta p}$=0.05$a$
which is sufficiently minor so as to still confer an overall benefit
to the device performance. Also shown in ~\figref{figure3}a is the
absorptivity spectra of just the GZO layer for the $\overline{\Delta
  p}$=0.05$a$ device (computed in FDTD again with three separate
simulations by subtracting the absorptivity of the device with GZO
absorption from one without which though not exact is still mostly
accurate) demonstrating the relatively small absorption in this very
thin 40nm layer that tends to increase somewhat with wavelength as the
absorption coefficient of $\mu$c-Si:H becomes smaller. The large
discrepancy between the experimental and simulated results at
wavelengths near 1000nm can be attributed to the weak sensitivity of
the detector in this region which is visible in the low
signal-to-noise ratio of the measured data. ~\Figref{figure3}b shows
the measured absorptivity spectra of the four PC samples and
demonstrates gradual broadening of the resonant-absorption peaks with
increasing disorder and thus the onset of quasi resonances consistent
with our prior observations~\cite{Oskooi12}. Note however that the
absorptivity spectra for the $\overline{\Delta p}$=0.20$a$ device
shows no traces of the resonant modes indicating that the amount of
partial lattice disorder necessary to broaden the absorption peaks to
the maximal extent without destroying them altogether - the key
principle for creating quasi resonances with optimal light-trapping
characteristics~\cite{Oskooi12} - in the actual fabricated samples as
a result of the intrinsic imperfections is markedly less than that of
the ideal designs typically used in the simulations. The absorptivity
spectra of the same device with an unpatterned $\mu$c-Si:H layer is
also shown for reference with its broad Fabry-P\'{e}rot resonances
where in the absence of a light-trapping structure the absorptivity at
each wavelength over the entire interval is considerably less than
that of its partially-disordered PC counterparts. The corresponding
data from an FDTD simulation is also shown for comparison to highlight
both the presence and quantity of constant background absorption in
the experiments due to the incomplete collection of light within the
integrating sphere.

We continue to experimentally measure the angular dependence of the
absorptivity spectra of the perfect and three partially-disordered
thin-film PC devices in ~\figref{figure4} at four angles of incidence
- 10$^\circ$, 20$^\circ$, 30$^\circ$ and 40$^\circ$ - again for
incident $\mathcal{S}$-polarized light. The performance of each device
is quantified using its light-trapping efficiency averaged over the
four angles as well as its standard deviation to gauge
robustness. Though the efficiency is calculated using the broadband
absorption of the entire device which includes that of the GZO layer,
the results are still relevant for assessing the performance of a
potential PV cell primarily because the GZO absorption is relatively
small as was demonstrated earlier. ~\Figref{figure4}a shows just how
sensitively the position of the resonant-absorption peaks of the
unperturbed PC structure depend on the incident conditions and the
degree to which this can be made less and less so with the gradual
addition of disorder just as we had previously observed with
simulations~\cite{Oskooi12}. The light-trapping efficiency increases
with partial lattice disorder (attaining a maximum of 56.8\% at
$\overline{\Delta p}$=0.05$a$ in ~\figref{figure4}b) before decreasing
considerably to 51.4\% at $\overline{\Delta p}$=0.20$a$ again
generally consistent with our earlier findings~\cite{Oskooi12}. The
dual integration of lattice disorder and fabrication imperfections is
perhaps best represented by the nearly-optimal partially-disordered PC
device with $\overline{\Delta p}$=0.10$a$ (in ~\figref{figure4}c)
which is nearly 10\% above in efficiency and more than twice as robust
as the ideal design (47.9\% $\pm$ 8.2\%) as computed by simulations.

In summary, we have experimentally shown how quasi resonances can be
incorporated into a thin-film silicon device as a PC with a controlled
amount of partial lattice disorder augmented with unintentional
fabrication-induced imperfections to boost light trapping over a broad
bandwidth and angular range. The design process was enabled by
accurate FDTD simulations to both structure the active-layer PC and to
more fully investigate absorption in the rest of the device. A further
extension of this design strategy to a complete thin-film
micro-crystalline PV cell possibly based on the guidelines we have
outlined is the next step for investigating the role of quasi
resonances on increasing overall efficiency.

\emph{This work was supported by Core Research for Evolutional Science
  and Technology (CREST) from the Japan Science and Technology
  agency. A.O. was supported by a postdoctoral fellowship from the
  Japan Society for the Promotion of Science (JSPS). We also thank
  Yosuke Kawamoto and Pedro Favuzzi of Kyoto University for fruitful
  discussions.}


\begin{thebibliography}{10}
\newcommand{\enquote}[1]{``#1''}

\bibitem{Poortmans06}
J.~Poortmans and V.~Arkhipov, \emph{Thin Film Solar Cells: Fabrication,
  Characterization and Applications} (John Wiley \& Sons, Ltd, 2006).

\bibitem{Pillai07}
S.~Pillai, K.~Catchpole, T.~Trupke, and M.~Green, \enquote{Surface plasmon
  enhanced silicon solar cells,} J.~Appl. Phys. \textbf{101} (2007).

\bibitem{Garnett10}
E.~Garnett and P.~Yang, \enquote{Light trapping in silicon nanowire solar
  cells,} Nano Lett. \textbf{10}, 1082--1087 (2010).

\bibitem{Han10}
S.~Han and G.~Chen, \enquote{Toward the lambertian limit of light trapping in
  thin nanostructured silicon solar cells,} Nano Lett. \textbf{10}, 4692--4696
  (2010).

\bibitem{Sheng11}
X.~Sheng, J.~Liu, I.~Kozinsky, A.~Agrawal, J.~Michel, and L.~Kimerling,
  \enquote{Design and non-lithographic fabrication of light trapping structures
  for thin film silicon solar cells,} Adv. Mater. \textbf{23}, 843--847 (2011).

\bibitem{Hsu12}
C.-M. Hsu, C.~Battaglia, C.~Pahud, Z.~Ruan, F.-J. Haug, S.~Fan, C.~Ballif, and
  Y.~Cui, \enquote{High-efficiency amorphous silicon solar cell on a periodic
  nanocone back reflector,} Adv. Ene. Mater. \textbf{2}, 628--633 (2012).

\bibitem{Sai12}
H.~Sai, K.~Saito, and M.~Kondo, \enquote{Enhanced photocurrent and conversion
  efficiency in thin-film microcrystalline silicon solar cells using
  periodically textured back reflectors with hexagonal dimple arrays,} Appl.
  Phys. Lett.  (2012).

\bibitem{Oskooi12}
A.~Oskooi, P.~Favuzzi, Y.~Tanaka, H.~Shigeta, Y.~Kawakami, and S.~Noda,
  \enquote{Partially-disordered photonic-crystal thin films for enhanced and
  robust photovoltaics,} Appl. Phys. Lett. \textbf{100} (2012).

\bibitem{Oskooi13}
A.~Oskooi, Y.~Tanaka, and S.~Noda, \enquote{Tandem photonic-crystal thin films
  surpassing lambertian light-trapping limit over broad bandwidth and angular
  range,} arxiv: 1304.5329  (2013).

\bibitem{Oskooi10}
A.~F. Oskooi, D.~Roundy, M.~Ibanescu, P.~Bermel, J.~D. Joannopoulos, and S.~G.
  Johnson, \enquote{{MEEP}: A flexible free-software package for
  electromagnetic simulations by the {FDTD} method,} Computer Physics
  Communications \textbf{181}, 687--702 (2010).

\bibitem{Python09}
M.~Python, O.~Madani, D.~Domine, F.~Meillaud, E.~Vallat-Sauvain, and C.~Ballif,
  \enquote{Influence of the substrate geometrical parameters on
  microcrystalline silicon growth for thin-film solar cells,} Sol. Ener. Mat.
  and Sol. Cells \textbf{93}, 1714--1720 (2009).

\bibitem{Palanchoke12}
U.~Palanchoke, V.~Jovanov, H.~Kurz, P.~Obermeyer, H.~Stiebig, and D.~Knipp,
  \enquote{Plasmonic effects in amorphous silicon thin film solar cells with
  metal back contacts,} Opt. Express \textbf{20}, 6340--6347 (2012).

\bibitem{Vallat06}
E.~Vallat-Sauvain, A.~Shah, and J.~Bailat, \emph{Advances in Microcrystalline
  Silicon Solar Cell Technologies} (John Wiley \& Sons, Ltd., 2006), chap.~4.

\bibitem{Vetterl00}
O.~Vetterl, F.~Finger, R.~Carius, P.~Hapke, L.~Houben, O.~Kluth, A.~Lambertz,
  A.~Muck, B.~Rech, and H.~Wagner, \enquote{Intrinsic microcrystalline silicon:
  A new material for photovoltaics,} Sol. Ener. Mat. \& Sol. Cells \textbf{62},
  97--108 (2000).

\bibitem{Green08}
M.~Green, \enquote{Self-consistent optical parameters of intrinsic silicon at
  300 k including temperature coefficients,} Solar Energy Materials and Solar
  Cells \textbf{92}, 1305--1310 (2008).

\bibitem{Yeh88}
P.~Yeh, \emph{Optical Waves in Layered Media} (Wiley, New York, 1988).

\end{thebibliography}
\end{document}